\documentstyle[11pt,adassconf]{article}  
\begin{document}  
\paperID{p2-53}
\title{Starfinder: a code for crowded stellar fields analysis}
\author{E.\ Diolaiti and O.\ Bendinelli}
\affil{Dipartimento di Astronomia, Universit\'a di Bologna,
       Via Ranzani,1, 40127 Bologna Italy, Email: diolaiti@bo.astro.it}
\author{D.\ Bonaccini, L.\ Close and D.\ Currie}
\affil{European Southern Observatory,Karl-Schwarzschild-Str.2, 
       D-85748 Garching b. Munchen, Germany }
\author{G.\ Parmeggiani}
\affil{Osservatorio Astronomico di Bologna, Via Ranzani, 1, 40127 Bologna, 
       Italy}
\contact{Emiliano Diolaiti}
\email{diolaiti@bo.astro.it}
\paindex{Diolaiti,E.}
\aindex{Bendinelli,O. }
\aindex{Bonaccini,C.}
\aindex{Close,L.}
\aindex{Currie,D.}
\aindex{Parmeggiani,G.}
\begin{abstract}    
Starfinder is an IDL code for the deep analysis of 
stellar fields, designed for 
well-sampled images with high and low Strehl factor. The Point Spread Function 
(PSF) for the analysis is extracted directly from the CCD frame, to take into 
account the actual structure of the instrumental response and the atmospheric 
effects. An important feature is represented by the possibility to measure the 
anisoplanatic effect in wide-field Adaptive Optics (AO) observations and 
exploit this knowledge to improve the analysis of the observed field. 
A description of the method and applications to real AO data are
presented.
\end{abstract}
\section{Introduction}
If the PSF is constant across the frame, the observed stellar field may 
be considered a superposition of shifted scaled replicas of the PSF itself, 
lying on a smooth background originated by faint unresolved stars and other 
possible sources.
 Actually the PSF is not always constant: in wide-field AO 
imaging, for instance, off-axis stars appear blurred and radially elongated 
with respect to the guide star; this anisoplanatic effect is mainly due to the partial 
correction of the wavefront tip-tilt. A further complication in the analysis of 
nearly diffraction-limited images is represented by the detailed structure of 
the PSF, which is generally difficult to model analytically and may produce 
false detections.
 Starfinder (see also Diolaiti et al, 1998) seeks to consider 
all these  aspects.
\section{ Analysis procedure}
\subsection{ PSF and background determination}
If it is not known, the PSF for the analysis must be extracted from the image.
In our code 
the user selects a set of stars, which are cleaned from the most contaminating
 sources, background-subtracted, 
centered with sub-pixel accuracy, 
normalized and superposed with a median operation. The halo of the retrieved 
PSF 
is then smoothed, applying a variable box size median filtering technique.

The PSF estimate represents a template to analyze the field stars; sub-pixel positioning
 is accomplished by interpolating the PSF array,  which must be well-sampled. 
A similar approach has been described in V\'eran et al. (1998). 

To overcome anisoplanatic effects in AO imaging we 
use an approximation of the local PSF given by the convolution  
of the reference source, commonly referred to as guide star, with a 
radially elongated elliptical gaussian. The parameters of this blurring kernel
are derived from a polynomial fit. To do this, first a set of stars at various 
distances from the reference 
source is 
selected  then the parameters (elongation and width) of the 
convolving elliptical gaussian, which gives the best match to the observation, are
 determined for each one . 
This set of measurements is fitted with a polynomial, 
which will be used to determine the local PSF for the analysis of each 
presumed star in the field.

The image background is estimated by interpolating a set of local measurements
 relative to sub-regions arranged in a regular grid (see Bertin et al, 1996). If the brightest
stars in the field can be removed, a very similar estimate may be obtained
by the application of a median smoothing technique to the input frame.
\subsection{Stars detection, astrometry and photometry}
The starting point is a list 
of presumed stars, whose observed intensity  in the 
background-removed image is greater than a prefixed detection threshold.
Preliminary smoothing
reduces the incidence of noise spikes.
The objects are listed by decreasing intensity and analyzed one by one
by the following sequence of steps:
\begin{enumerate}
\item re-identification after subtraction of the known stars, in order to reject 
spurious detections due to PSF features of bright sources;
\item cross-correlation with the PSF, as a measure of similarity with 
the template;
\item astrometric and photometric analysis by local fitting.
\end {enumerate}
  Each new accepted star is added to a 
synthetic stellar field, updated at every step. 
When all the objects in the 
list 
have been analyzed, a final re-fitting is performed to improve their astrometry 
and photometry; then they are temporarily removed to upgrade the background 
estimate.

The basic step described above (detection and analysis) may be repeated: a new 
list of presumed stars is formed after subtracting the previously detected ones 
and the analysis is started again on the original image. This iteration is very 
useful to detect stars in crowded groups, down to separations comparable to the 
Rayleigh limit for the detection of close binaries.

An optional deblending mode is available. All the objects somewhat more extended
 than the PSF are considered 
blends. The deblending strategy consists of an iterative 
search for residuals around the object and subsequent fitting; the iteration 
stops when no more residual is found or the fit of the last residual was not 
successful.
\section{ Applications to high and low Strehl images}
The algorithm has been run on a K-band PUEO image of the Galactic Center, 
as an example of a well-sampled high-Strehl AO observation of a 
stellar field. 
\begin{figure}
\epsscale{0.6}
\plottwo{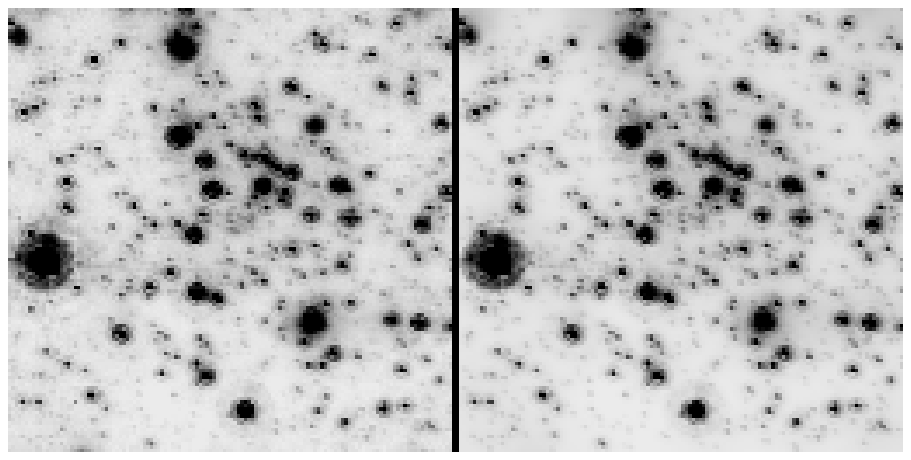}{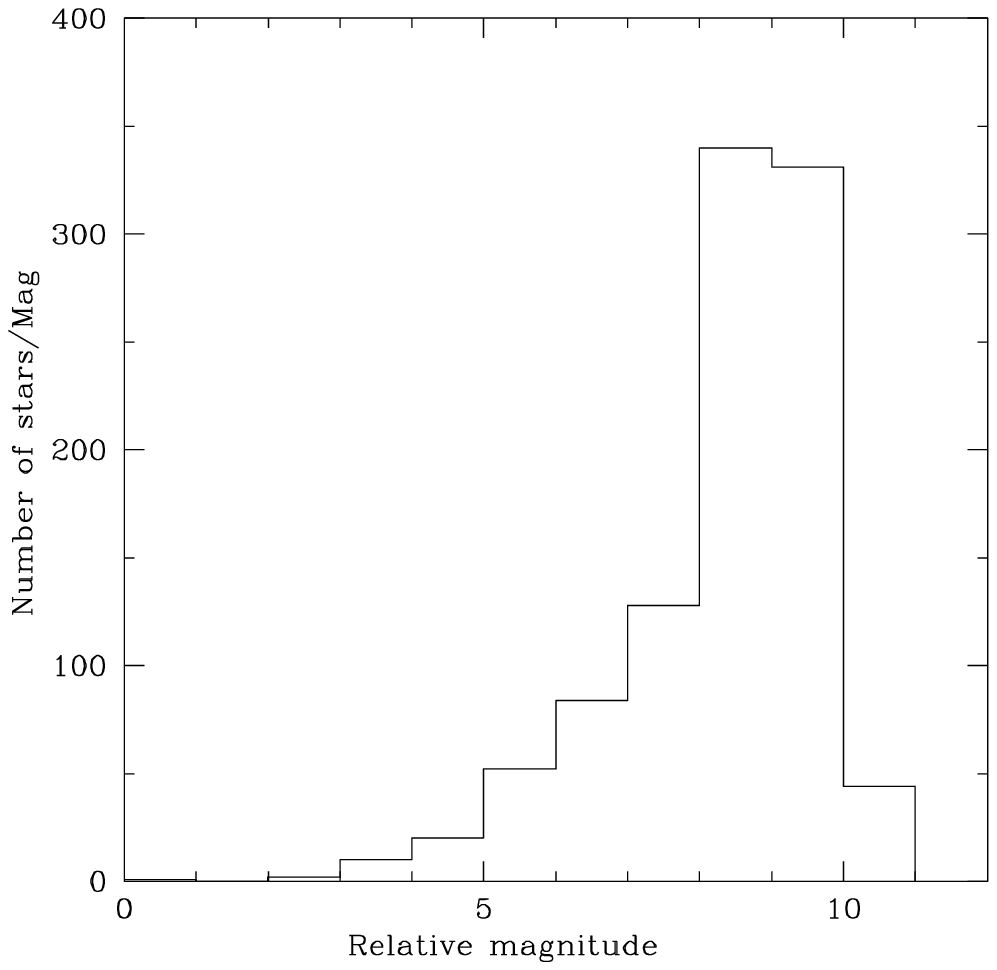}
\caption{Left to right. PUEO image of the Galactic Center;
 reconstructed image 
given by the sum
 of the synthetic stellar field (more than 1000 stars) 
and the estimated background,  
the display stretch is square root; the obtained luminosity function}
 \end{figure} 
We have evaluated the astrometric and photometric accuracy of the algorithm  adding to the image,
for each magnitude bin in the retrieved luminosity function (fig.1),  
a total of 10 \%  of  
synthetic stars located at random positions, with the only constraint 
that the minimum distance of each simulated star from all the previously 
detected ones must be greater than 1 PSF FWHM (about 4 pixels).
The plot of the photometric errors 
shows accurate and unbiased photometry (see fig.2a,2b). 
\begin{figure}
\epsscale{0.5}
\plotone{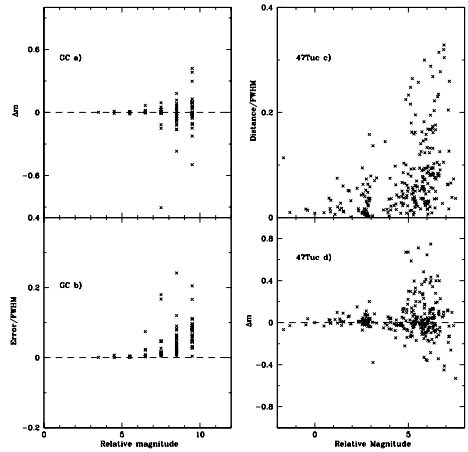}
\caption{Galactic Centre: a)  
photometric 
errors of detected synthetic stars, b) astrometric errors, representing 
the distance 
between the true and the calculated position in FWHM units. 47 Tuc: c) off-centering between
corresponding stars in the two images, d) magnitude difference. The first 4 
points correspond to repaired saturated stars}
\end{figure}

Our method has also been applied to two well-sampled low-Strehl images of the 
globular cluster 47~Tuc, observed at the ESO 3.6m telescope with the ADONIS AO 
system.  The PSF 
FWHM is about 6 pixels.The two frames, that have a large overlap area, have been
analyzed with the same 
procedure.  The results 
(fig. 2 c and d)
present a good  internal astrometric and 
photometric accuracy.
\section{ Conclusions and future developments}
Starfinder seems to be able to analyze well-sampled images of very crowded 
fields observed, for instance, with ground-based AO systems. According to our
experience it may be successfully applied also to adequately sampled HST images, 
like those obtained with dithering strategies.
It is reasonably fast (only few minutes
on a Pentium II PC for the analysis of the Galactic Center) and a 
widget interface makes it accessible to users unfamiliar with IDL.
In the near future the tools for space-variant analysis will be improved.
A further work will give a complete description of the code.
\acknowledgments
Francois Rigaut is acknowledged for kindly providing the PUEO image of the 
Galactic Center and for supporting the initial development of this method.
%

%

\end{document}